# Social Group Query Based on Multi-fuzzy-constrained Strong Simulation

Guliu Liu[1,2], Lei Li[1,2,*], Senior *Member, IEEE*, Guanfeng Liu[3], *Member, IEEE*, and Xindong Wu[1,4,2], *Fellow, IEEE*

[1](Key Laboratory of Knowledge Engineering with Big Data (Hefei University of Technology), Ministry of Education)

[2](School of Computer Science and Information Engineering, Hefei University of Technology)

[3](Department of Computing, Macquarie University)

[4](Mininglamp Academy of Sciences, Mininglamp Technology)

Traditional social group analysis mostly uses interaction models, event models, or other methods to identify and distinguish groups. This type of method can divide social participants into different groups based on their geographic location, social relationships, and/or related events. However, in some applications, it is necessary to make more specific restrictions on the members and the interaction between members of the group. Generally, graph pattern matching (GPM) is used to solve this problem. However, the existing GPM methods rarely consider the rich contextual information of nodes and edges to measure the credibility between members. In this paper, a social group query problem that needs to consider the trust between members of the group is proposed. To solve this problem, we propose a Strong Simulation GPM algorithm (NTSS) based on the exploration of pattern Node Topological ordered sequence. Aiming at the inefficiency of the NTSS algorithm when matching pattern graph with multiple nodes with zero in-degree and the problem of repeated calculation of matched edges shared by multiple matching subgraphs, two optimization strategies are proposed. Finally, we conduct verification experiments on the effectiveness and efficiency of the NTSS algorithm and the algorithms with the optimization strategies on four social network datasets in real applications. Experimental results show that the NTSS algorithm is significantly better than the existing multi-constrained GPM algorithm, and the NTSS_Inv_EdgC algorithm, which combines two optimization strategies, greatly improves the efficiency of the NTSS algorithm.

## 1 INTRODUCTION

The query and identification of groups are of great significance in the analysis and application of social networks, such as the discovery of social groups [1,2, 3], the identification of criminal groups [4], and the positioning of expert groups [5]. The social network-based group query can define the pattern graph of the group to be queried by restricting the members and the relationship between the members in the group.

In 2013, Fan et al. [5] proposed the problem of graph pattern matching (GPM) for expert group matching in collaborative networks. For example, in the collaborative network shown in Fig. 1(b), each participant has three attributes: name, job, and work experience. The edges between them represent cooperative relationships. If a company wants to form a team for system development, the team needs system architect, business analyst, project manager, system developers, and testers to work together. The job requirements, experience requirements, the relationship between the team members, and the path length requirements are shown in Fig. 1(a). Obviously, the influence of the relationship between team members on the efficiency of team collaboration is not considered. For example, if BA lacks communication with PM and SD, or cannot communicate well with PM and SD, it will affect the development efficiency of the team. Besides, the professional ability of each person is also an important factor to be considered for teamwork. For a team, the credibility of members and the trust between members have a very important impact on the formation of the team, and the measurement of trust between members are affected by many factors. Therefore, we propose a social group query problem that needs to measure the credibility among team members.

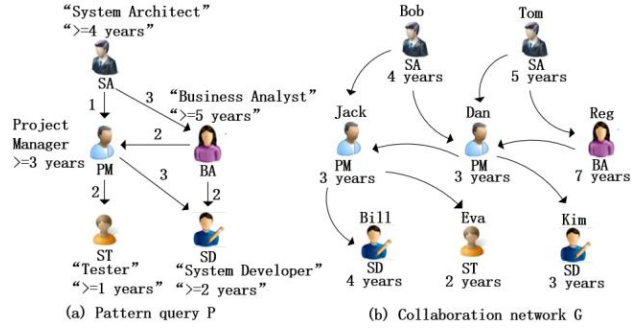

Fig. 1. Query pattern graph $P$ and collaboration network $G$

In 2015, Liu et al. [6] considered crowdsourced tourism and social network-based e-commerce in contextual social networks may require multiple constraints on matching nodes and edges, so they proposed the problem of Multi-Constrained Graph Pattern Matching (MC-GPM) and proposed Multi-Constrained Simulation (MCS) matching method. MC-GPM can be well adapted to the group query problem. However, the Baseline algorithm and the HAMC algorithm proposed by Liu et al. [6], and the M-HAMC algorithm [7] proposed later, are essentially similar to the exact GPM algorithms, and both adopt exploration-based approaches. Although those algorithms can return matching results faster, it is difficult to complete the task of completely matching large-scale social network data graphs. In addition, MCS cannot well match the topological relationship between nodes in the pattern graph, so the matching results will contain a large number of matching results that do not meet the requirements.

In this paper, we focus on the social group queries that require the calculation of credibility among members. For each social group to be queried, there is generally a leader or an organizer. In this paper, we assume that the pattern graph to be queried is a directed acyclic graph starting from the leader (or organizer). Since the existing MC-GPM algorithms are similar to exact GPM algorithms, each leader may appear in multiple matching subgraphs, which can be considered as one large matching subgraph for group queries. In addition, previous definitions of MC-GPM only added judgment of multiple constraints on nodes and edges based on bounded simulation [4], so it could not well match the topology of pattern graphs just like bounded simulation.

In response to the above problems, the MC-GPM matching model is improved firstly, and the Multi-Fuzzy-Constrained Strong Simulation (MFCSS) matching model is proposed. MFCSS ensures that the matched nodes completely satisfy the topology of the corresponding pattern node through the dual simulation [10]; by limiting the locality of the entire matching subgraph, it further ensures that the matching subgraph composed of matched nodes has a more similar topological structure to the pattern graph. Then, given the high time complexity of the existing MC-GPM algorithm, a Strong Simulation matching algorithm (NTSS) based on the exploration of pattern Node Topological ordered sequence is proposed. NTSS ensures that the matched nodes and the corresponding pattern node have the same topological structure by examining the predecessor and subsequent topological structure of the matching candidate nodes. Furthermore, the exploration-based method of NTSS ensures the locality of the matching subgraph. Then, in view of the low matching efficiency of the NTSS algorithm when matching pattern graphs with multiple vertices with zero in-degree (i.e., multi-source pattern graphs), an optimization algorithm NTSS_Inv for reverse edge matching is proposed. Aiming at the problem that the NTSS algorithm needs to repeatedly calculate the shared matching paths (or edges) of the matching subgraphs, an optimization algorithm NTSS_EdgC for matched paths caching is proposed. Finally, we implement the NTSS, NTSS_Inv, NTSS_EdgC, and the NTSS_Inv_EdgC algorithm with two optimization strategies, and conduct confirmatory experiments on four social network datasets of different sizes. The experimental results show that the proposed NTSS algorithm is significantly better than the existing MC-GPM algorithms, and the NTSS_Inv_EdgC algorithm greatly improves the efficiency of the NTSS algorithm.

## 2 RELATED WORK

Regarding the discovery and recognition of groups, Ngan et al. [11] used the characteristics of the recognition of people in the picture and the characteristics of crowd gathering to group the characters; Tran et al. [12] identified people in communication through modeling of participant interaction clues in social networks; Amin et al. [13] used event-based models to compare groups in multiple networks to discover criminal groups; Yang et al. [14,15] analyzed the time, location and social relationship of the characters in social networks to organize social activities. This kind of method is not close to our work, so we will not discuss it in detail.

The earliest GPM refers to the matching subgraph [16] with the same topology structure as the pattern graph, so it is also called subgraph matching, subgraph isomorphism, or GPM based on isomorphism. This kind of matching is often used for biological data analysis [17,18] and social network applications with strict requirements on the structure of subgraphs [19], and often use indexing [20,21] and pruning [22] methods to improve the matching efficiency of the algorithm. For some emerging applications in social networks, isomorphism-based matching requirements are too strict, and the efficiency of the algorithm is too low. In 2010, Fan et al. [4] proposed bounded simulation matching based on node simulation to solve the query problem of the crime group. The algorithm can complete the GPM task within cubic time. Since bounded simulation matching only constrains the successor topology of candidate nodes, the matching results include a large number of matched nodes that do not meet the requirements of the predecessor topology of the pattern nodes. In response to this problem, Ma et al. [10] proposed strong simulation matching. Strong simulation matching requires that the matched node must meet the dual simulation requirements, that is, the matched node have to not only meet the subsequent topology of the corresponding pattern node but also meet the predecessor topology of the pattern node. In addition, strong simulation matching further requires the matched nodes of a matching subgraph to be in a subgraph with a diameter of D (D=2d, d represents the diameter of the pattern graph), that is, the matching subgraph is required to meet locality, thus further ensuring that the topology structure of the matching subgraph is consistent with that of the pattern graph.

The above work does not consider the information on the edges of the social network graph, that is, the constraints of the information about the relationship between nodes. In 2015, Liu et al. [6] first proposed the MC-GPM problem in contextual social networks and proposed the heuristic algorithm HAMC. Since the HAMC algorithm cannot fully match large-scale social network data, Liu et al. further studied the method of returning top-k matching results [23] and the parallel M-HAMC algorithm [7]. Li et al. [8] studied the application of MC-GPM in system reliability, and regarded it as a multi-objective constraint problem, and used multi-objective optimization methods to filter the GPM results to return more valuable results, but they did not improve the MC-GPM algorithm itself. Considering that there may be problems with fuzzy attributes in large-scale graph data, Liu et al. [9] proposed a general problem model for GPM in big graph data, multi-fuzzy-constrained graph pattern matching (MFC-GPM), and proposed a method ETOF-K based on exploration of edge topological order in the pattern graph. This algorithm improves the efficiency of the MC-GPM by adjusting the edge matching order and optimizing the matching method of multi-constrained pattern edges, but it is still an algorithm similar to exact GPM in essence, and its time complexity is very high.

The edge matching of the MFC-GPM problem is a multi-constrained optimal path selection (MCOPs) problem. For two social participants A and B (there is no direct correlation between them), there may be multiple paths between them. In the case of multiple constraints, finding the optimal path is proved to be an NP-Complete problem [24]. In 2010, Liu et al. [25] proposed the problem of calculating the optimal trust path between nodes in complex social networks and proposed to use the weighted sum of trust and intimacy between nodes and the influence of nodes in the field of social networks to evaluate whether B is trustworthy to A. In 2012, Liu et al. [26] proposed a calculation model for a trusted network in a complex contextual social network, adding the similarity of the hobbies between nodes and the length of the relationship path into the calculation model for comprehensive trust measurement, and through evaluation of the trusted network formed by multiple paths between A and B to measure the credibility of B to A.

## 3 GRAPH PATTERN MATCHING

This part first gives general definition of data graph, pattern graph, and matching subgraph, and then introduces the proposed multi-fuzzy-constrained strong simulation (MFCSS).

### 3.1 Related Terms

GPM means that given a pattern graph P and a data graph G, query all subgraphs of the matching pattern graph P from G. The definitions of the data graph, the pattern graph, and the matching subgraph are as follows.

A data graph is a directed graph with node and edge attributes, which can be denoted as $G = (V, E, f_v^D, f_e^D)$, where:
- $V$ represents a set of nodes;
- $E$ represents the set of edges, $(v_i, v_j) \in E$ represents the directed edge from node $v_i$ to node $v_j$;
- $f_v^D$ represents a function defined on the node set $V$, $\forall v \in V$, $f_v^D(v)$ represents the attributes set of nodes $v$;
- $f_e^D$ represents a function defined on the edge set $E$, $\forall (v_i, v_j) \in E$, $f_e^D(v_i, v_j)$ represents the attributes set on the edge $(v_i, v_j)$.

A pattern graph is a directed graph with attributes constraints on nodes and edges. It can be expressed as $P = (V_P, E_P, f_v^P, f_e^P, f_l^P, f_m^P)$, where:
- $V_P$ and $E_P$ represent the set of pattern node and the set of pattern edge, respectively;
- $f_v^P$ represents the function defined on the pattern node set $V_P$; $\forall (u_i, u_j) \in E_P$, $f_e^P(u_i, u_j)$ represents the attributes constraints on the pattern node $u$;

- $f_e^P$ represents the function defined on the pattern edge set $E_P$, $\forall (u_i, u_j) \in E_P, f_e^P(u_i, u_j)$ represents the attributes constraints defined on the pattern edge $(u_i, u_j)$;
- $f_l^P$ represents the function defined on the pattern edge set $E_P$, $\forall (u_i, u_j) \in E_P, f_l^P(u_i, u_j)$ represents the matching path length constraint defined on the pattern edge $(u_i, u_j)$;
- $f_m^P$ represents a set of membership constraint functions and its corresponding set of membership constraint values. For each attribute constraint in the pattern graph, a membership function can be defined to calculate whether the attribute (or aggregated attribute) on the matched node or matched edge (or path) meets the membership constraint, and the corresponding membership constraint value can be set.

The matching subgraph is a subgraph that matches the pattern graph in the data graph $G$, denoted as $G_{sub}=(V_{sub}, E_{sub}, f_{v_{sub}}^D, f_{e_{sub}}^D)$, where $G_{sub} \subset G, V_{sub} \subset V, E_{sub} \subset E, f_{v_{sub}}^D \subset f_v^D, f_{e_{sub}}^D \subset f_e^D$.

### 3.2 Multi-fuzzy-constrained Strong Simulation

The MCS is the same as the bounded simulation, that is, it cannot guarantee that the topological structure of the pattern graph is matched, nor can it guarantee the connectivity of the matching subgraphs. Fan et al. [10] proposed strong simulation matching, which requires that the matching subgraph must meet dual simulation, and the nodes in the matching subgraph must be in the subgraph with $v$ as the center node and radius $r$; thereby ensuring the matched nodes meet the topological requirements of the pattern nodes and the connectivity of the matching subgraph. However, strong simulation is edge-to-edge matching and does not consider multiple constraints on the pattern edges. Therefore, this paper proposes MFCSS on the basis of MFC-GPM. This section will first introduce the meaning of some related concepts, and then give the definition of MFCSS.

About the meaning of the concepts of Dual Simulation, $dist(v, v')$, radius $r$, diameter $D$, etc., Fan et al. have been given in the paper [10], and will not be repeated in this paper. This paper focuses on the fuzzy multi-constrained social group query problem that needs to measure the credibility between members. Regarding the measurement of credibility, the method proposed by Liu et al. in [6, 25] through the aggregation calculation of the trust value, social intimacy, and social influence factor on the connection path between participants is used. However, in this paper, we have given a more reasonable explanation for these attributes. In addition, we also introduced the aggregation attributes and the calculation methods of the aggregation attributes and explained the form of the matching subgraph of the social group query study. The specific introduction of related concepts is as follows:

**Social influence factor:** The social influence factor is determined by the number of recognized remarks and the number of recognized experiences or experiences in a certain domain. The more remarks and experience recognized by others, the higher the social influence factor of the participant in the field can be considered. The social influence factor is represented by $\rho$, and $\rho \in [0,1]$.

**Social trust:** Social trust refers to the personal scores given to each other's credibility by two adjacent participants in a social network based on their previous communication experience, denoted by $T$, and $T \in [0,1]$.

**Social intimacy:** Social intimacy is determined by the type of relationship between two social participants, such as relatives, teachers and students, friends, etc., as well as the frequency of communication between them. The closer the relationship and the more frequent the interaction, the higher the social intimacy. Social intimacy is represented by $R$, and $R \in [0,1]$.

**Aggregation attribute:** For two social participants A and B, suppose the path from A to B is represented as A, $v_1$, $v_2$, ... $v_m$, B, where m is a positive integer greater than or equal to 1. Aggregation attribute refers to the aggregate value of an attribute obtained by aggregation, averaging and other methods for the attributes on the nodes or edges on the path. In this paper, the method of averaging is used for the aggregation of social influence factors on the matched path, while the method of integration based on the trust propagation model [27] is used for social trust and social intimacy.

**Matching subgraph:** Given a node $v \in V$, $v$ matches the leader in the pattern graph, with this node as the central node and the diameter $D$ of the pattern graph as the radius $r$, can get a data subgraph $\hat{G}[v, r] \subset G$. Then the subgraph composed of all matched nodes obtained from $v$ as the starting point and the nodes and edges on the matched paths in $\hat{G}[v, r]$ is a matching subgraph of the pattern graph.

**Multi-fuzzy-constrained Strong Simulation:** Given a data graph $G = (V, E, f_v^D, f_e^D)$ and a pattern graph $P = (V_P, E_P, f_v^P, f_e^P, f_l^P, f_m^P)$, $G$ matches $P$ through multi-fuzzy-constrained strong simulation means that there is such a binary relationship $S \subset V \times V_P$, which satisfies:

- $(u, v) \subset S$, if there is a membership function for node attributes in $f_m^P$, the corresponding attributes in $f_v^D(v)$ only need to satisfy the corresponding membership constraints, otherwise $f_v^D(v)$ need to satisfy the constraints $f_v^P(u)$ defined on the node $u$;
- $\forall u \in V_P, \exists (u, v) \subset S, v \in V$;
- $\forall (u, v) \subset S$, for all $(u, u') \in E_P$, there exists $path(v, v') \in G$, $(u', v') \subset S$, at the same time for all $(u'', u) \in E_P$, there exists $path(v'', v) \in G$, $(u'', v'') \subset S$;

for all matched paths $path(v, v')$ (here $path(v, v')$ generally refers to $path(v, v')$ and $path(v'', v)$ ), $length(path(v, v')) \leq f_l^P(u, u')$, and if there is a membership function for the aggregated attribute on the matched path in $f_m^P$, then the corresponding aggregated attribute in $path(v, v')$ only need to meet the corresponding membership constraints; otherwise, the aggregated attributes on the matched path need to meet the corresponding attribute constraint in $f_e^P$; all nodes $v$ in $S$ are contained in a data subgraph with $v_s$ as the central node and radius $r$, where $r$ is equal to the diameter $D$ of the pattern graph.

**Example 1:** Consider a social group query problem that needs to calculate the credibility between members of the group. The data graph can be denoted as $G = (V, E, f_v^D, f_e^D)$, where $f_v^D$ represents the label and social influence factor $\rho$ of $v$, $f_e^D$ represents social trust $T$ and social intimacy $R$ between participants. The pattern graph can be denoted as $P = (V_P, E_P, f_v^P, f_e^P, f_l^P, f_m^P)$, where $f_v^P$ represents the label constraint on the pattern node and the social influence factor constraint $\rho_v$, $f_e^P$ represents the social trust constraint $\lambda_T$, social intimacy constraint $\lambda_R$, and social influence factor constraint $\lambda_\rho$ on the pattern edge, $f_l^P$ represents the matching path length constraint on the pattern edge, $f_m^P = \{f_{\rho_v}^m, f_T^m, f_R^m, f_\rho^m, \rho_{vm}, T_m, R_m, \rho_m\}$, where $f_{\rho_v}^m$ represents the membership function defined on the social influence factor constraint $\rho_v$, and $\rho_{vm}$ represents the corresponding membership constraint value. $f_T^m, f_R^m, f_\rho^m$ respectively represent the membership functions defined on the pattern edge attribute constraints $\lambda_T, \lambda_R, \lambda_\rho$, and $T_m, R_m$, and $\rho_m$ respectively represent the membership constraint values of the corresponding attribute constraints. The pattern graph $P$ matches the data graph $G$ through multi-fuzzy-constrained strong simulation, which means that there is a binary relationship $S \subset V \times V_P$, which satisfies:

- if $(u, v) \subset S$, then $f_v^D(v)$ satisfies the label constraint defined on the node u and $f_{\rho_v}^m(\rho(v)) \geq \rho_{vm}$.
- $\forall u \in V_P, \exists (u, v) \subset S, v \in V$;
- $\forall (u, v) \subset S$, for all $(u, u') \in E_P$, there exists $path(v, v') \in G$, $(u', v') \subset S$, and for all $(u'', u) \in E_P$, there exists $path(v'', v) \in G$, $(u'', v'') \subset S$;

for all matched paths $path(v, v')$ (here $path(v, v')$ refers to $path(v, v')$ and $path(v'', v)$), $len(path(v, v')) \leq f_l^P(u, u')$, and $f_T^m(AT^{D_i}(v, v')) \geq T_m$, $f_R^m(AR^{D_i}(v, v')) \geq R_m$, $f_\rho^m(A\rho^{D_i}(v, v')) \geq \rho_m$, where $AT^{D_i}(v, v')$, $AR^{D_i}(v, v')$, $A\rho^{D_i}(v, v')$ represent the aggregated attributes value of $T, R, \rho$ on $path(v, v')$, respectively. for all nodes $v$ in $S$, they

---
Method 1: $getCandidate$
---

Input: $u \in V_P, V \in G$
Output: $Cand_u$
Begin
1.       while there is a $v \in V$ not visited do
2.           if $label_v(u) \subset label_v(v)$ and $f_{\rho_v}^m(\rho(v)) \geq \rho_{vm}$
3.              add $v$ to $Cand_u$
4.       return $Cand_u$
End

---
ALGORITHM 1: NTSS
---

Input: $G, P$
Output: $G_{sub}^{All}$
Begin
1.       Get $V_T$ and $V_E$;
2.       For $V_T[0]$, call $getCandidate(u)$ to get $Cand_u$;
3.       $i = 0$;
4.       while $i < length(Cand_u)$ do
5.           $v_s = Cand_u[i]$;
6.           $G_{sub} = TopologicalMatching(v_s, V_T, V_E, G, P)$;
7.           if $G_{sub} \neq \emptyset$
8.              Add $G_{sub}$ to $G_{sub}^{All}$;
9.           $i = i + 1$;
10.     return $G_{sub}^{All}$;
End

are contained in a subgraph with $v_s$ as the center node and the diameter of the pattern graph P as the radius, where $v_s$ refers to the matching starting node in the pattern graph, and he/she is generally the leader of the group.

## 4 NTSS ALGORITHM

The execution process of the existing MC-GPM algorithm is generally divided into two parts, the matching of multi-constrained pattern edges and the connection of matched paths based on the topology of the pattern graph. Generally, the execution of simulation-based GPM algorithms is divided into two parts: getting the candidate nodes-sets and filtering the candidate nodes based on the topology of the pattern nodes. This difference is because, for MC-GPM, the matching of multi-constrained pattern edges is much more difficult than the general simulation-based GPM. It is difficult to return the matching result in a short time using the existing simulation-based algorithms, so the existing MC-GPM algorithms all use exploration-based methods to achieve the requirement of quickly returning matching results. However, the matching algorithm based on exploration, due to the locality of the matching process, it is difficult to avoid the problem of repeated calculation of common matching paths between matching subgraphs. Moreover, the existing MC-GPM algorithms do not design matching algorithms strictly according to the idea of multi-constrained simulation but adopt the method similar to exact subgraph matching, so there is also the problem of subgraph explosion.

In response to the above problems, we first propose a strong simulation matching algorithm (NTSS) based on the exploration of pattern node topological ordered sequence, which is significantly better than the existing MFC-

<pre>
                                  ALGORITHM 2: TopologicalMatching
</pre>

Input: $v_s, V_T, V_E, G, P$
Output: $G_{temp}^v$
Begin
1.      For all $(u, u') \in V_P, (u, v_s) \in S$
2.          $pLM(v_s, v') = edgeMatch(v_s, (u, u'))$;
3.          $G_{temp}^v = findSuitablePath(pLM(v_s, v'))$;
4.          if $SuitP(v_s, v') = \emptyset$
5.              return $\emptyset$;
6.      $i = 1$;
7.      while $i < length(V_T)$ do
8.          $u_c = V_T[i]$;
9.          if $u_c \in V_E$
10.             call $getCandidate(u_c)$ to get $Cand_{u_c}$;
11.         else
12.             get $Cand_{u_c}$ from $G_{temp}^v$;
13.         $j = 0$;
14.         while $j < length(Cand_{u_c})$ do
15.             $v_c = Cand_{u_c}[j]$;
16.             if $inEdgeCheck(v_c) == false$
17.                 $j = j + 1$;
18.                 $deleteCandidate(v_c)$;
19.                 continue;
20.             For all $(u, u') \in V_P, (u_c, v_c) \in S$
21.                 $pLM(v_c, v_c') = edgeMatch(v_c, (u_c, u_c'))$;
22.                 $G_{temp}^v = findSuitablePath(pLM(v_c, v_c'))$;
23.                 if $SuitP(v_c, v_c') = \emptyset$
24.                     $deleteCandidate(v_c)$;
25.             $j = j + 1$;
26.         $i = i + 1$;
27.     return $G_{temp}^v$;
End

GPM algorithms in efficiency. Then we add two optimizations to the NTSS algorithm and propose the NTSS_Inv_EdgC algorithm, which greatly reduces the time complexity and improves the efficiency of the NTSS algorithm. This section introduces the NTSS algorithm in detail, and the next section will introduce the optimization methods for NTSS.

### 4.1 Node Topological Based Pattern Node Matching

The matching of pattern nodes can be divided into two stages. In the first stage, candidate nodes are selected according to the constraints on the pattern nodes in the pattern graph; in the second stage, candidate nodes are filtered according to the topology structure of the corresponding pattern nodes.

In the social group query problem described in this paper, constraints on nodes include constraint on node label $label_v$ and constraint on node social influence factor $\rho_v$. In addition, $f_m^P$ also contains a membership function $f_{\rho_v}^m$ for the node social influence factor constraint $\rho_v$ and the corresponding membership constraint value $\rho_{vm}$. For a pattern node $u$, if there is a node $v \in V$ is a candidate node for $u$, then there are $label_v(u) \subset label_v(v)$ and $f_{\rho_v}^m(\rho(v)) \geq \rho_{vm}$. The candidate node set $Cand_u$ of $u$ can be obtained through the $getCandidate$ method, as shown in Method 1.

The NTSS algorithm matches pattern nodes according to the topologically ordered sequence. Because, when considering whether a node satisfies the predecessor and successor topological relationship, based on the topologically ordered matching order, it can directly determine whether the candidate node satisfies the predecessor topology. If it is not satisfied, it is no longer necessary to judge whether the subsequent topology meets the requirements, so as to achieve the purpose of pruning. In addition, we assume that there is a leader (or organizer) in the group, the matching of the pattern graph starts from the leader, and it is a pattern node with zero entry. The execution steps of the NTSS algorithm are shown in ALGORITHM 1.

The input of the algorithm is the data graph $G$ and the pattern graph $P$, and the output is the set of matching subgraphs $G_{sub}^{All}$. The algorithm first uses the topological sorting algorithm to obtain the topologically ordered sequence $V_T$ of the pattern nodes, and at the same time records the out-degree $out_u$ and in-degree $in_u$ of each node. Then, obtain the pattern node-set $V_E$ which in-degree is zero. The above steps are shown in lines 1-2 of ALGORITHM 1. Loop through the candidate set of the starting node, call the $TopologicalMatching$ method to obtain the matching subgraph organized or lead by $v = Cand_u[i]$. If the matching result is not an empty set, then added the matching subgraph $G_{sub}$ to the matching subgraph set $G_{sub}^{All}$, as shown in lines 4-9 of ALGORITHM 1.

The specific execution steps of the $TopologicalMatching$ method is shown in ALGORITHM 2. First, use the $edgeMatch$ and $findSuitablePath$ methods, match all matching paths $SuitP(v_s, v')$ that satisfy the multiple constraints of the pattern edge $(u, u') \in V_P$, where $v_s$ matching start node of pattern graph, $(u, v_s) \in S$, and add these matched paths to the data structure $G_{temp}^v$ that stores the intermediate result of the matching subgraph, as shown in ALGORITHM 2 lines 1-3. The storage structure of $G_{temp}^v$ is shown in Fig. 2. It is an array of key-value pairs, each element in the array is a key-value pair, the key is the pattern node $u_i$ ($0 \leq i < n_p$, $n_p$ represents the number of nodes in the pattern graph), and the value is the matching candidate node list of $u_i$. $v_{i,j}$ ($0 \leq i < n_p, 1 \leq j \leq m_i$) represents the $j$th matched node of the $i$-th pattern node $u_i$, and $m_i$ represents the number of matching candidate nodes of the $i$-th pattern node $u_i$. For each matched node $v_{i,j}$, its storage structure in $G_{temp}^v$ is shown in Fig. 3. Let $v$ represent $v_{i,j}$ in Fig. 2, $listN$ represents the list of predecessor nodes of pattern node $u$ ($(u,v) \in S$), and $listE$ represents the list of successor edges of pattern node $u$ ($(u,v) \in S$). When using the edge matching function $findSuitablePath$ to get a matching path $SuitP(v, v_k')$, it needs to be added to the matching list of the $k$th subsequent edge of $listE$, and at the same time in the pattern node $u_k'$, $(u_k', v_k') \in S$ add $v_k'$ to the list of matching candidate nodes, and then find the precursor node $u$, $(u, u_k') \in E_P$ in the $listN$ of the matching candidate node $v_k'$, and add $v$ to the matching list of $u$. Therefore, after matching the starting node's subsequent edges, not only can we get the matched path lists of all the starting node's subsequent pattern edges $(u, u') \in V_P$, but also get the candidate nodes set $v'$ of $u'$ based on the candidate node $v$, and stored in $G_{temp}^v$.

$TopologicalMatching$ method, then, starting from the second node of $V_T$, the pattern nodes are matched sequentially in topological order. When the current pattern node to be matched $u_c = V_T[i]$ is the successor of the matched pattern node, obtain the candidate node set of $u_c$ from $G_{temp}^v$, otherwise, call the method $getCandidate$ to obtain the candidate node set of $u_c$, as in ALGORITHM 2 shown in lines 9-12. After obtaining the candidate node set $Cand_{u_c}$, loop through the elements $v_c$ in the candidate node set. First, determine whether $v_c$ meets the topological requirements of the incoming edges, that is, determine whether there is a null value in the matched node lists of all the precursor nodes in the *listN* of $v_c$. If it is not satisfied, the *deleteCandidate* method is called, and the matching candidate nodes that do not meet the topology requirements are recursively deleted, and then the next candidate node is looped; if it is satisfied, the *edgeMatch* and *findSuitablePath* methods are called to obtain the matching paths of all subsequent edges $(u_c, u_c') \in E_P$, and adds them to $G_{temp}^v$, as shown in ALGORITHM 2 lines 13-25.

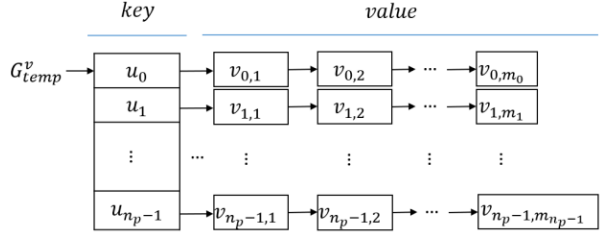

Fig. 2. The data structure of intermediate result $G_{temp}^v$ of matching subgraph $G_{sub}$

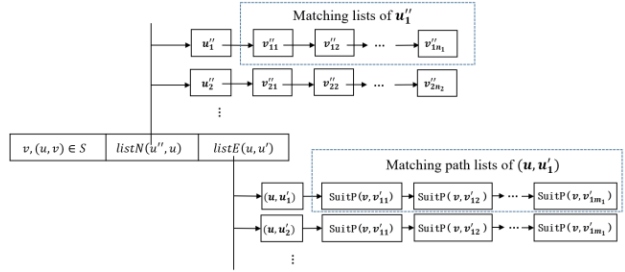

Fig. 3. Match data node $v$ structure in $G_{temp}^v$

The *edgeMatch* and *findSuitablePath* methods called in the *TopologicalMatching* method are both part of multi-constrained pattern edge matching methods, which will be introduced in detail in section IV B. The *deleteCandidate* method is a method to recursively delete matching candidate nodes and edges in $G_{temp}^v$ that do not meet the topology requirements.

The NTSS algorithm is an exploration-based simulation matching algorithm. It can be adapted to different applications according to different edge matching algorithms, such as edge-to-edge matching, edge-to-path matching, multi-constrained edge matching etc. At the same time, NTSS is an algorithm that satisfies strong simulation matching, that is, NTSS has the following two properties.

**Proposition 1:** NTSS meets dual simulation. $\forall v, (u, v) \in S$, then $\forall (u, u') \in E_P$, there must exists $path(v, v')$ matching it, where $(u', v') \in S$. At the same time, $\forall (u'', u) \in E_P$, there must exists $path(v'', v)$ to match with it, where $(u'', v'') \in S$.

**Proof:** For any matching candidate node $v, (u, v) \in S$, *inEdgeCheck* must be executed. If there is $(u, u') \in E_P$ and there is no $path(v, v')$ matching it, then $v$ will be added to *delCS* and then deleted from $G_{temp}^v$, which contradicts with $(u, v) \in S$. At the same time, $\forall v, (u, v) \in S$, if there is $(u'', u) \in E_P$, there is no $path(v'', v)$ matching it, then if $v$ is the starting node $v_s$, the matching subgraph $G_{temp}^v$ for $v_s$ will return an empty set, that is $(u, v_s) \notin S$; if v is not the starting node, then $v$ will be added to *delCS*, and then deleted from $G_{temp}^v$, which contradicts with $(u, v) \in S$. The proposition is proved.

ALGORITHM 3: *edgeMatch*

Input: $v$, $(u, u')$, $G$
Output: $pLM(v, v')$
Begin
1.     $Q = \emptyset$;
2.     $Q.push(pathE)$;
3.     while $Q.empty() == false$ do
4.       $pathj(v, v') = Q.front()$;
5.       $Q.pop()$;
6.       if $pathj(v, v')$ not in $MPL^v_{(u,u')}$
7.         if there is another path from $v$ to $v'$ in $MPL^v_{(u,u')}$
8.           add $pathj(v, v')$ to the $pathlist(v, v')$;
9.         else
10.           add $pathj(v, v')$ to $MPL^v_{(u,u')}$ directly;
11.       get adjacency set $adjL_v$ of $v$;
12.       while $adjL_v \neq \emptyset$ do
13.         $adj_v = getOneElement(adjL_v)$;
14.         $remove(adj_v, adjL_v)$;
15.         if $len(pathj(v, v')) < boundedlen(u, u')-1$
16.           $pathi = pathj(v, v') + (v, adj_v)$;
17.           $Q.push(pathi)$;
18.         else if $len(pathj(v, v')) == boundedlen(u, u')-1$
19.           if $adj_v$ satisfies the constraints of $u'$
20.             $pathi = pathj(v, v') + (v, adj_v)$;
21.             $Q.push(pathi)$;
22.     return $pLM(v, v')$
End

**Proposition 2:** NTSS satisfies locality. That is, for any matching subgraph $G_{sub}$, any matched node $v$, $(u, v) \in S$, all $dist(v_s, v)$ less than or equal to $D$, and $D$ represents the pattern graph diameter.

**Proof:** Suppose $(u_s, v_s) \in S$, $u_s$ represents the leader in the pattern graph, that is, the starting node of pattern matching, and $v_s$ is the matched node. It is easy to know that $dist(u_s, u) \leq D$, where $u$ represents any pattern node in the pattern graph. Use $u_s, u_1, u_2, \ldots, u_n, u$ to represent the shortest path from $u_s$ to $u$, then $dist(u_s, u) = length(u_s, u_1) + length(u_1, u_2) + \cdots + length(u_n, u)$. $\forall v, (u, v) \in S$, there must exists a path $v_s, v_1, v_2, \ldots, v_n, v$ from $v_s$ to $v$, where $v_1, v_2, \ldots, v_n$ match $u_1, u_2, \ldots, u_n$, respectively, so there are $length(v_i, v_{i+1}) \leq length(u_i, u_{i+1})$. Thus, $dist(v_s, v) \leq dist(u_s, u) \leq D$, the proposition is proved.

### 4.2 Multi-fuzzy-constrained Pattern Edge Matching

The multi-fuzzy-constrained pattern edge matching is a MCOPs problem, which is an NP-complete problem [24]. To reduce the calculation of unnecessary aggregated attributes and improve the efficiency of edge matching, this paper adopts a pattern edge matching method that firstly searches the path that satisfies the constraint length, and then aggregates the constraint attribute values on the path. The pathfinding method that satisfies the constraint length is shown in ALGORITHM 3.

The overall idea of ALGORITHM 3 is to start from node $v$, breadth-first traverse the edges in the data graph, and record each traversal path $pathj(v, v')$ starting from $v$. When the end point $v'$ of the path matches $u'$, Add $pathj(v, v')$ to the matched path set $pLM(v, v')$ that satisfies the path length constraint. The $pathE$ in ALGORITHM 3 represents an empty path whose start and end nodes are both $v$. To reduce unnecessary enqueue and dequeue operation when the path length of $pathj(v, v')$ is equal to the constraint length of $(u, u')$ minus one, we first judge whether $adj_v$ matches $u'$, and then execute the enqueue operation as shown in lines 18-21 of ALGORITHM 3. Among the multiple matched paths from $v$, there may be multiple $v'$ that satisfy the constraints defined on the pattern node $u'$, and for the same $v'$, there may also be multiple paths from $v$ to $v'$. In order to facilitate the selection of the optimal matched path in the next stage, for each $v'$ that matches $u'$, we build a path list $pathlist(v, v')$ from $v$ to $v'$, as shown in lines 6-10 of ALGORITHM 3.

After obtaining all matching paths that satisfy $boundedlen(u, u')$ starting from $v$, we need to call the $findSuitablePath$ method to review the multiple constraints defined on the pattern edge $(u, u')$. For the matching of multi-fuzzy-constrained pattern edge, the specific steps are shown in Algorithm 4. For the path list $pathlist(v, v')$ from $v$ to $v'$, calculate the aggregate attributes value on each path $AT^{D_i}(v, v'), AR^{D_i}(v, v'), A\rho^{D_i}(v, v')$, and record the smallest aggregate attribute value $min\_attr_c$ on each path, compare $min\_attr_c$ of all paths, and select the path with the largest $min\_attr_c$ as the optimal matched path from $v$ to $v'$, and then the optimal matched path is added to the intermediate result $G_{temp}^v$ of the matching subgraph.

For the convenience of calculation, the membership functions $f_{\rho_v}^m, f_T^m, f_R^m, f_\rho^m$ of each attribute mentioned in this paper all take the form shown in formula 1, where $AP_x$ represents the attribute (aggregation attribute) of the matched node (path), $\lambda_x$ represents the constraint of the corresponding attribute. The membership constraint value $\rho_{vm}, T_m, R_m, \rho_m$ of each attribute are all set to 0.9.

$$f = \begin{cases} \frac{AP_x}{\lambda_x} & AP_x < \lambda_x \\ 1 & AP_x \geq \lambda_x \end{cases} \quad (1)$$

### 4.3 Complexity Analysis

The NTSS algorithm needs to traverse all nodes in the data graph to obtain the candidate node set of the starting node $u_s$, and the time complexity is $O(n)$. For each candidate node $v_s$ of the starting node $u_s$, an exploratory MFC-GPM process based on the topology order of the pattern nodes is required. For the nodes in the candidate node set of each pattern node, it is necessary to judge whether the predecessor topology structure is satisfied and the subsequent edges are matched. Use $M$ to represent the average out-degree of nodes in the pattern graph, and using $O(edge)$ to represent the time complexity of multi-fuzzy-constrained edge matching, then the time complexity of the NTSS algorithm is $O(n * n_P * n * M * edge)$.

For the matching of each pattern edge, first, it is necessary to query the path from $v$ to multiple $v'$ ($v'$ matches $u'$) with a path length not exceeding $boundedlen(u, u')$. That is, you need to perform a breadth-first path traversal with a depth of $boundedlen(u, u')$ (abbreviated as $l$) starting from $v$. Then it is necessary to judge multiple constraints for each matched path that already satisfies the path length constraint. Use $N$ to represent the average out-degree of nodes in the data graph, and the time complexity of multi-fuzzy-constrained edge matching is $O(2 * N^l)$.

Therefore, the overall time complexity of the NTSS algorithm is $O(2n^2 n_P M N^l)$, which is $O(n^2 n_P M N^l)$ after removing the constant term.

## 5  OPTIMIZATION TECHNIQUES

The NTSS algorithm proposed in this paper still uses an exploration-based method in order to quickly obtain the matching results, but the node matching and the multi-fuzzy-constrained edge matching are optimized, which improves the efficiency of the algorithm and makes it reach the strong simulation matching requirements. Even so, the NTSS algorithm still exists some problems. This section proposes optimization methods for the inefficiency of the NTSS algorithm in processing multi-source pattern graphs and the problem of repeated calculation of some common matching paths between matching subgraphs. The optimization methods in these two cases are respectively introduced in the following sections.

### 5.1  Reverse Edge Matching

Consider the pattern graph shown in Fig. 4(a). Assuming PM is the group leader in the pattern graph, the topologically ordered sequence of nodes in the pattern graph is PM, BA, SD, ST. When ALGORITHM 2 completes the successor edge matching of a candidate node of PM, the $getCandidate$ method needs to be called to obtain the candidate node set $Cand_{BA}$ of BA. When the data graph is large, there may be tens of thousands or even more candidate nodes of BA, and there may be only a small part of them that can link to the same SD node as the current matched nodes of PM. For GPM starting from each candidate node of PM, the $getCandidate$ method needs to be called to retrieve the set $Cand_{BA}$ of BA, and then perform subsequent edge matching and connection with the current SD candidate node for each candidate node in $Cand_{BA}$ one by one. It can be seen from this that the NTSS algorithm has very low matching efficiency for multi-source pattern graphs, and there are a lot of unnecessary and repeated calculations. Since the time complexity of executing the $getCandidate$ method is $O(n)$, and the time complexity of executing BA's subsequent edge matching is $O(MN^l)$, the time complexity of NTSS algorithm is $O(n^{(m+2)} n_P M^{(m+1)} N^{(m+1)l})$ when there are $m$ nodes with zero indegree in the pattern graph.

To reduce the time complexity of NTSS algorithm matching multi-source pattern graphs, a reverse edge matching method is proposed, which can effectively reduce unnecessary edge matching calculations. The specific method of improvement is that if the current pattern node $u_c$ has no matching candidate nodes, and there is a set of candidate nodes of the successor node $u'_c$ of the current pattern node $u_c$ in $G^v_{temp}$, we start from each candidate node of $u'_c$ and use the inverse adjacency list of the data graph to reversely match the pattern edges $(u_c, u'_c)$ to obtain the candidate node set $Cand_{u_c}$ of $u_c$. Then it traverses the candidate node set of $u_c$ to match other subsequent edges. With the improved NTSS algorithm (NTSS_Inv), the time complexity is restored to $O(n^2 n_P M N^l)$ when matching multi-source pattern graphs.

### 5.2  Cache Mechanism of Matched Paths

Consider the pattern graph shown in Fig. 4(a). When the length constraint of the matching path from BA to SD is changed to 3, another additional matched $path$ (Reg, Kim) of the pattern edge (BA, SD) can be obtained. Thus, the NTSS algorithm can be used to obtain two matching subgraphs. The matched nodes set of the first matching subgraph is $V_{sub1}$ ={Bill, Reg, Elly, Kim, Eva}, and the matched nodes set of the second matching subgraph is $V_{sub2}$ ={Bob, Reg, Elly, Kim, Eva}. It can be seen that there are multiple identical matched nodes in the two matching subgraphs, and it is also easy to know that the two matching subgraphs have multiple identical matched paths. But when matching two subgraphs, these same matched paths need to be recalculated. In addition, consider there is a graphical designer (GD) between BA and SD in Fig. 4(a), we still assuming that PM is the leader of the group, and the topologically ordered sequence of pattern nodes in the graph is PM, BA, GD, SD, ST. When ALGORITHM 2 is executed to complete the subsequent edge matching of a candidate node of PM, there is neither a candidate node of BA nor a candidate node of GD in $G_{temp}^v$, so the $getCandidate$ method must be called to obtain the candidate nodes set $Cand_{BA}$ of node BA. Through the analysis in section V A, it is easy to know that when using the NTSS algorithm to deal with this kind of problem, there are a large number of repeated calculation problems about the matching of BA and its subsequent edges.

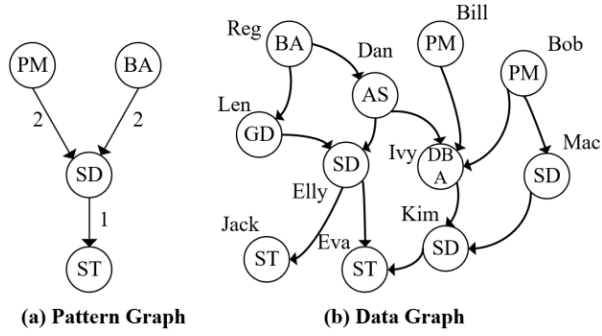

(a) Pattern Graph    (b) Data Graph

Fig. 4. Pattern graph and data graph of bounded simulation

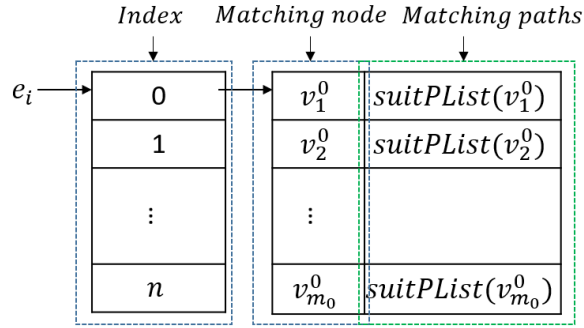

Fig. 5. Cache structure of matched edges

In response to those problems, this section proposes a method of caching the matched paths, thereby avoiding the problem of repeated calculation of the matched paths in the NTSS algorithm. First, the edges in the pattern graph are numbered, and use $e_i$ to represent the pattern edge numbered $i$. Then construct the data structure shown in Fig. 5 to store the edge matching results. The first column represents the pattern edge numbered $i$, the second column represents the list of nodes that match the starting node of the pattern edge numbered $i$, and the third column represents the list of matched paths matching the pattern edge numbered $i$ obtained from each matched node. When using the $TopologicalMatching$ algorithm in the NTSS algorithm to match the subsequent edges of the matching candidate node $v$, it can first query the cache shown in Fig. 5 whether the matching of the current node $v$ on its subsequent edges exists in the cache. Because in the NTSS algorithm that based on matched paths cache (NTSS_EdgC), the subsequent edge matching of each matched node is only calculated once, and the sum of all matching candidate nodes is less than or equal to the sum of nodes in the data graph, so the time complexity of the NTSS_EdgC algorithm is $O(n*M*2*N^l)$, which is $O(nMN^l)$ after removing the constant term.

Table I: The Social Datasets

| Name | Vertices | Edges | Description |
|---|---|---|---|
| Epinions | 75879 | 508837 | A trust-oriented social network |
| Slashdot | 77360 | 905468 | A friend/foe social network |
| Pokec | 1632803 | 30622564 | General online social network |
| LiveJournal | 4847571 | 68993773 | General online social network |

Table II: Statistics of the Number of Matching Subgraphs

| Datasets | NTSS | | Fuzzy-ETOF-K | |
|---|---|---|---|---|
| | $P_1$ | $P_2$ | $P_1$ | $P_2$ |
| Epinions | 109 | 275 | 109 | 275 |
| Slashdot | 215 | 515 | 214 | 515 |
| Pokec | 1588 | 30 | 1583 | 30 |
| LiveJournal | 8101 | 265 | 8097 | 265 |

## 6 EXPERIMENTAL STUDY

To verify the superiority of the NTSS algorithm proposed in this paper and the effectiveness of the two optimization strategies proposed for the NTSS algorithm, the following comparative experiments are carried out in this section.

### 6.1 Experimental Implements and Settings

We used four social network datasets of different sizes as the experimental datasets. The detailed statistics of the datasets are shown in Table I. These datasets only provide the association between the social participants but do not provide the attributes such as the social influence factor of the nodes, the social intimacy and trust of the association relationship required by our experiment. These attributes can be obtained by data mining methods and are not within the scope of this paper. In order not to lose generality, we used python to preprocess the datasets and use the $random$ function to randomly generate those attribute values on the nodes and edges, and the range of those attribute values is 0 to 1.

The pattern graphs and parameter settings used in our experiment is shown in Fig. 6. Because the Fuzzy-ETOF-K algorithm and the NTSS algorithm have high time complexity when processing multi-source pattern graphs. when 0.5 is used as the constraints condition of nodes and edges in pattern graph (b) in Fig.9, the matching task cannot be completed within a reasonable time in Pokec and LiveJournal datasets. Therefore, in the experiment, when experimenting on the Pokec and LiveJournal datasets, the constraint values on the nodes and edges are set to 0.8 for the pattern graph (b).

About the implementation of the algorithm, we first obtained the source code of the existing MFC-GPM algorithm Fuzzy-ETOF-K and then programmed the NTSS algorithm by ourselves. Aiming at the two optimization strategies proposed in this paper, we have implemented three versions of optimization algorithms of NTSS: NTSS_Inv, NTSS_EdgC, and NTSS_Inv_EdgC. All the above algorithms are completed using C++ based on Visual Studio 2015. The experiments in this paper are all running on a PC with Intel(R) Core (TM) i7-8700K CPU @3.70 GHz, 48 GB RAM, Windows 10 operating system.

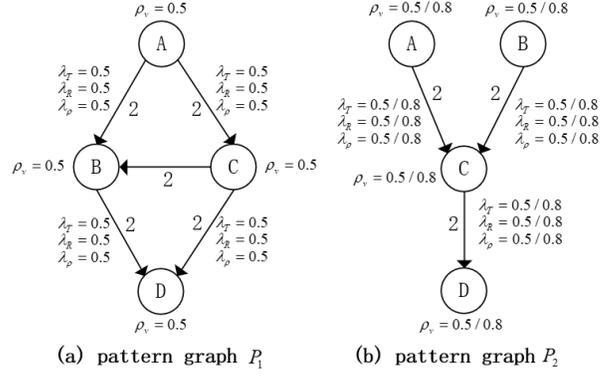

Fig. 6. Pattern graphs of experimental

### 6.2 Experimental Results and Analysis

**Exp-1: Effectiveness.** This experiment verifies the effectiveness of the NTSS algorithm by comparing the number of matching subgraphs returned by the NTSS algorithm and the Fuzzy-ETOF-K algorithm, as well as the average number of matched nodes and matched paths contained in each matching subgraph. When counting the number of matching subgraphs of the Fuzzy-ETOF-K algorithm, since the Fuzzy-ETOF-K algorithm is similar to exact subgraph matching, there is a problem that the same matching starting node returns multiple matching results. Therefore, before the comparison, we aggregated the matching results with the same starting matched node and regarded them as a matching result.

The statistics of the number of matching subgraphs returned by the NTSS algorithm and the Fuzzy-ETOF-K algorithm are shown in Table II, and the average statistics of the number of matched nodes and matched paths contained in each matching subgraph are shown in Fig. 7. $P1$ and $P2$ marked in the legend represent the matching results of the corresponding algorithm when matching the pattern graphs $P_1$ and $P_2$ in Fig. 6 respectively.

It can be seen from Table II that the NTSS algorithm and the Fuzzy-ETOF-K algorithm return the same number of results when matching the pattern graph $P_2$. When matching the pattern graph $P_1$, the number of matching results of the NTSS algorithm is slightly more than that of the Fuzzy-ETOF-K algorithm. At the same time, it can be seen from Fig. 7 that when the pattern graph $P_2$ is matched, the matching subgraph obtained by the two algorithms contains exactly the same number of matched nodes and paths. When matching the pattern graph $P_1$, the number of matched nodes and paths in the matching subgraph obtained by the NTSS algorithm is slightly more than that of the Fuzzy-ETOF-K algorithm. The reason for this is that the NTSS algorithm is an algorithm that satisfies the strong simulation matching, while the Fuzzy-ETOF-K algorithm is a similar precise matching algorithm. For the topological structure of the pattern graph and the data graph as shown in Fig. 8, the NTSS algorithm can be matched but the Fuzzy-ETOF-K algorithm cannot be matched. In summary, the effectiveness of the NTSS algorithm is proved.

**Exp-2: Efficiency.** This experiment is divided into three parts: The first part verifies the efficiency of the NTSS algorithm by comparing the number of matching subgraphs returned by the NTSS algorithm and the Fuzzy-ETOF-K algorithm over time. The second part verifies the effectiveness of the two optimization strategies and the efficiency of the NTSS_Inv_EdgC algorithm by comparing five algorithms: Fuzzy-ETOF-K, NTSS, NTSS_Inv, NTSS_EdgC, and NTSS_Inv_EdgC. In the third part, by further comparing the three algorithms of NTSS_Inv, NTSS_EdgC, and NTSS_Inv_EdgC, the stability and efficiency of the NTSS_Inv_EdgC algorithm are verified. The statistics of the total execution time of each algorithm are shown in Table III. The $k_1$ and $k_2$ in the table represent the ratio of the execution time of the Fuzzy-ETOF-K algorithm to the NTSS_Inv_EdgC algorithm and the ratio of the execution time of the NTSS algorithm to the NTSS_Inv_EdgC algorithm. The relationship between the number of matching results returned by each algorithm and the time consumed is shown in Fig. 9 to Fig. 12.

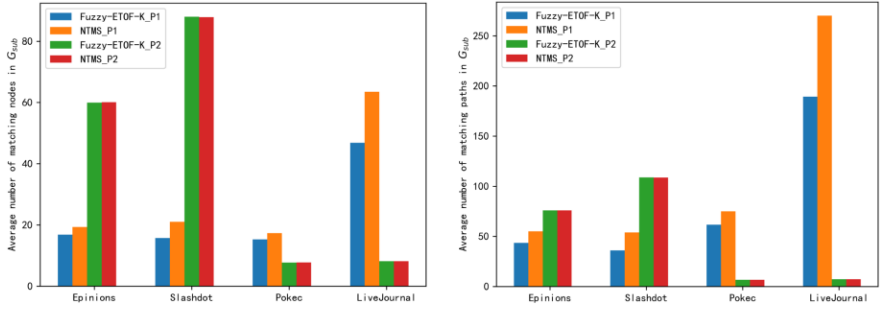

Fig. 7. The mean of the number of matched nodes and paths contained in matching subgraphs

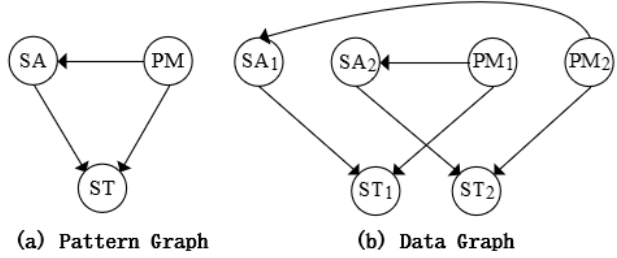

Fig. 8. The pattern graph and data graph are used to illustrate the difference between the NTSS algorithm and the Fuzzy-ETOF-K algorithm

**Part 1: Verification of the efficiency of the NTSS algorithm**

The relationship between the execution time and the number of returned results when the NTSS algorithm and the Fuzzy-ETOF-K algorithm match the pattern graphs $P_1$ and $P_2$ in Fig. 6 is shown in Fig. 9 and Fig. 10, respectively. As can be seen from Fig. 9, when matching the pattern graph $P_1$, the number of results returned by the NTSS algorithm increases with the execution time significantly faster than the Fuzzy-ETOF-K algorithm. As can be seen from Fig. 10, when matching the pattern graph $P_2$, although the execution efficiency of the NTSS algorithm is obviously better than the Fuzzy-ETOF-K algorithm on the Epinions and Slashdot datasets, the execution efficiency on the Pokec and LiveJournal datasets is only a little better than Fuzzy-ETOF-K algorithm. This is because the time complexity of NTSS algorithm increases when processing multi-source pattern graphs. The larger the data scale, the greater the impact of this complexity change on the actual execution time. In general, although the efficiency of the NTSS algorithm decreases when matching multi-source pattern graphs, the efficiency of the NTSS algorithm is higher than the Fuzzy-ETOF-K algorithm when matching two pattern graphs.

**Part 2: Verification of the effectiveness of two optimization strategies**

It can be seen from Fig. 9 that the NTSS_EdgC and NTSS_Inv_EdgC algorithms that have added matched paths caching strategies are significantly better than the NTSS algorithm and the Fuzzy-ETOF-K algorithm. The execution efficiency of the NTSS_Inv algorithm with the reverse edge matching strategy is basically the same as the NTSS algorithm. This is because there are no pattern edges that need to be reversed matched in the pattern graph $P_1$, so the NTSS_Inv algorithm is theoretically equal to the efficiency of the NTSS algorithm. However, in the actual matching process, in the NTSS_Inv algorithm, we only query the candidate nodes once for the pattern nodes with zero entry degree, so the efficiency is slightly higher than that of the NTSS algorithm. As shown in Fig. 10, the efficiency of the NTSS algorithm with the optimization strategy is significantly better than the NTSS algorithm and the Fuzzy-ETOF-K algorithm. This is because when matching the pattern graph $P_2$, not only can reverse edge matching be used to reduce unnecessary candidate node search and matched path connections, but there are also a large number of matched paths recalculation that can be optimized by the matched paths caching strategy, so both optimization strategies can play a good role in improving the execution efficiency of the NTSS algorithm. In general, both optimization strategies can improve the efficiency of the NTSS algorithm, but the reverse edge matching can only improve the efficiency of the NTSS algorithm for matching multi-source pattern graphs. The cache strategy of the matched paths has better stability, and can improve the efficiency of the NTSS algorithm in both cases.

Table III: Statistics of the Number of Matching Subgraphs

| Datasets | P | Fuzzy-ETOF-K | NTSS | NTSS_Inv | NTSS_EdgC | NTSS_Inv_EdgC | $k_1$ | $k_2$ |
|---|---|---|---|---|---|---|---|---|
| Epinions | $P_1$ | 17.46 | 7.42 | 7.92 | 2.36 | **2.17** | 8.05 | 3.42 |
| | $P_2$ | 276.27 | 152.60 | 13.14 | 3.57 | **2.84** | 97.38 | 53.73 |
| Slashdot | $P_1$ | 107.68 | 25.50 | 27.57 | 6.12 | **5.66** | 19.01 | 4.51 |
| | $P_2$ | 1415.15 | 805.94 | 28.27 | 8.04 | **6.91** | 204.68 | 116.63 |
| Pokec | $P_1$ | 4940.94 | 221.12 | 216.55 | 82.70 | **77.66** | 63.62 | 2.85 |
| | $P_2$ | 9161.83 | 8968.24 | 13.82 | 28.24 | **12.82** | 714.82 | 699.55 |
| LiveJournal | $P_1$ | 22335.90 | 3758.60 | 3728.46 | 379.59 | **352.01** | 63.45 | 10.68 |
| | $P_2$ | 73404.30 | 70114.92 | 78.87 | 119.89 | **56.26** | 1304.71 | 1246.27 |

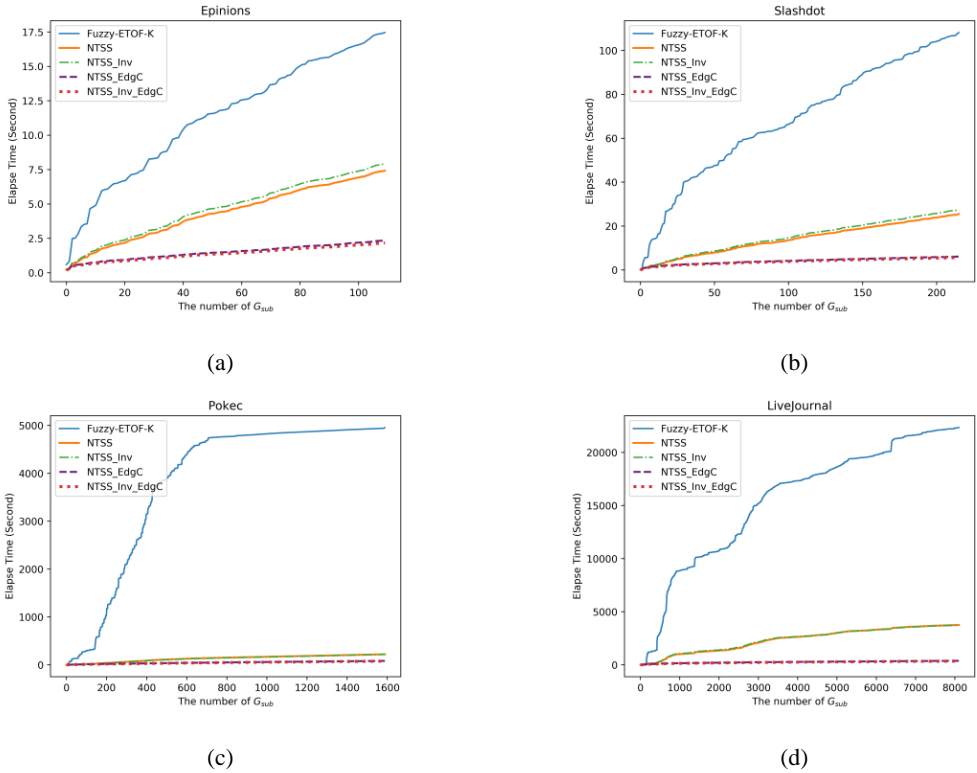

Fig. 9. The change in the number of results returned by all comparison algorithms over time when matching $P_1$

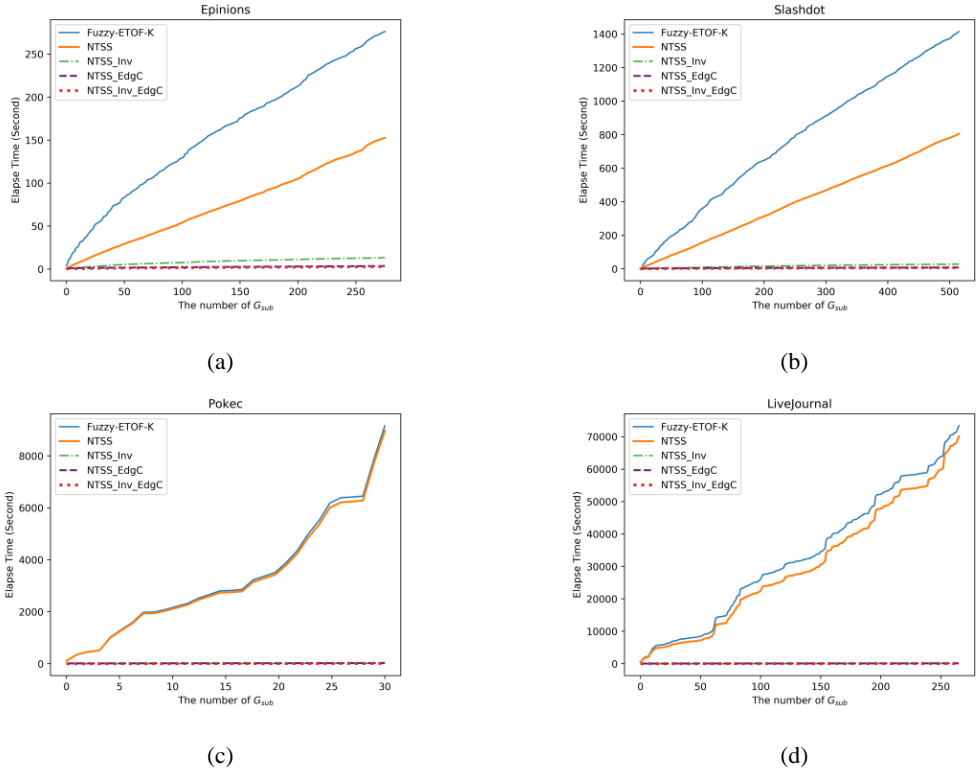

Fig. 10. The change in the number of results returned by all comparison algorithms over time when matching $P_2$

**Part 3: The stability and efficiency of the NTSS_Inv_EdgC algorithm**

In Fig. 9 and Fig. 10, it is difficult to observe the difference of the efficiency among NTSS_Inv, NTSS_EdgC and NTSS_Inv_EdgC, because the execute time of them is too little compared with Fuzzy-ETOF-K algorithm and part of the experiment of NTSS algorithm. Therefore, in this section, we further compare the time consumption of the three algorithms in matching pattern graphs $P_1$ and $P_2$. The experimental results are shown in Fig. 11 and Fig. 12 respectively.

As can be seen from Fig. 11, when matching the pattern graph $P_1$, the execution efficiency of the NTSS_Inv algorithm is far lower than that of the NTSS_EdgC and NTSS_Inv_EdgC algorithms, and the execution efficiency of the latter two algorithms are similar. The reason for this, as mentioned above, is because there are no pattern edges in $P_1$ that require reverse edge matching. It can be seen from Fig. 10 and Fig. 12 that although the two optimization strategies can play a very good role in optimizing the efficiency of the NTSS algorithm when matching $P_2$. However, on the Epinions and Slashdot datasets, the NTSS_EdgC algorithm is more efficient than the NTSS_Inv algorithm, and the NTSS_Inv algorithm on the Pokec and LiveJournal datasets is more efficient than the NTSS_EdgC algorithm.

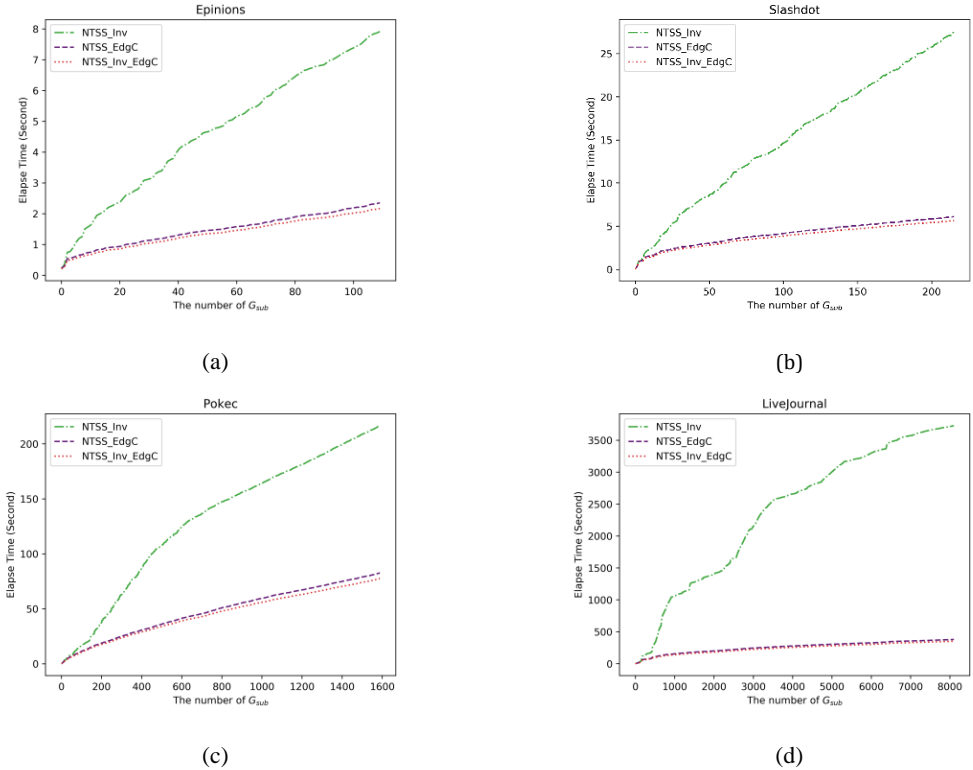

Fig. 11. The change in the number of results returned by the algorithm with optimization strategy over time when matching $P_1$

However, the NTSS_Inv_EdgC algorithm, which combines two optimization strategies, has the stability and best-matching efficiency when matching the two pattern graphs. In addition, as shown in Table III, when matching the pattern graph $P_1$, the consumption time of the Fuzzy-ETOF-K algorithm is 8-63 times of the NTSS_Inv_EdgC algorithm, and the NTSS algorithm is 2-10 times of the NTSS_Inv_EdgC algorithm. When matching the pattern graph $P_2$, the consumption time of the Fuzzy-ETOF-K algorithm is 97-1304 times of the NTSS_Inv_EdgC algorithm, and the NTSS algorithm is 53-1246 times of the NTSS_Inv_EdgC algorithm. This proves that the NTSS_Inv_EdgC algorithm has a significantly better matching efficiency than the Fuzzy-ETOF-K and NTSS algorithm.

**Exp-3: Memory Usage.** In this experiment, the memory usage of each algorithm for GPM on four different datasets was statistically compared. Fig. 13(a) and Fig. 13(b) show the statistical results of memory usage when matching the pattern graphs $P_1$ and $P_2$, respectively. It can be seen from the Fig. 13 that the memory usage of the same algorithm when matching two pattern graphs is basically the same. NTSS_Inv algorithm and NTSS_Inv_EdgC algorithm use nearly twice as much memory as the other three algorithms because they need to store the inverse adjacency list of the data graph when they perform reverse edge matching.

## 7 CONCLUSION

In this paper, we propose a social group query problem that needs to calculate the credibility between members in the group, MFCSS matching model is proposed. Then, to address the MFCSS matching, the NTSS algorithm is proposed. Aiming at reducing the high time complexity of the NTSS algorithm in matching multi-source pattern graphs, we propose an optimization algorithm NTSS_Inv that using a reverse edge matching strategy. To avoid the repeated search calculation for the common matching paths of multiple matching subgraphs in the matching process of the NTSS algorithm, an optimization algorithm NTSS_EdgC with matched paths cache is proposed. Finally, we implement the three algorithms: NTSS, NTSS_Inv, and NTSS_EdgC, and the NTSS_Inv_EdgC algorithm with two optimization strategies added, and carry out comparative experiments with the existing MFC-GPM algorithm Fuzzy-ETOF-K on four social network datasets. The experimental results show that the NTSS algorithm is better than the existing MC-GPM algorithm Fuzzy-ETOF-K, and the improved NTSS algorithm NTSS_Inv_EdgC significantly improves the efficiency of NTSS, and is many times more efficient than the Fuzzy-ETOF-K algorithm.

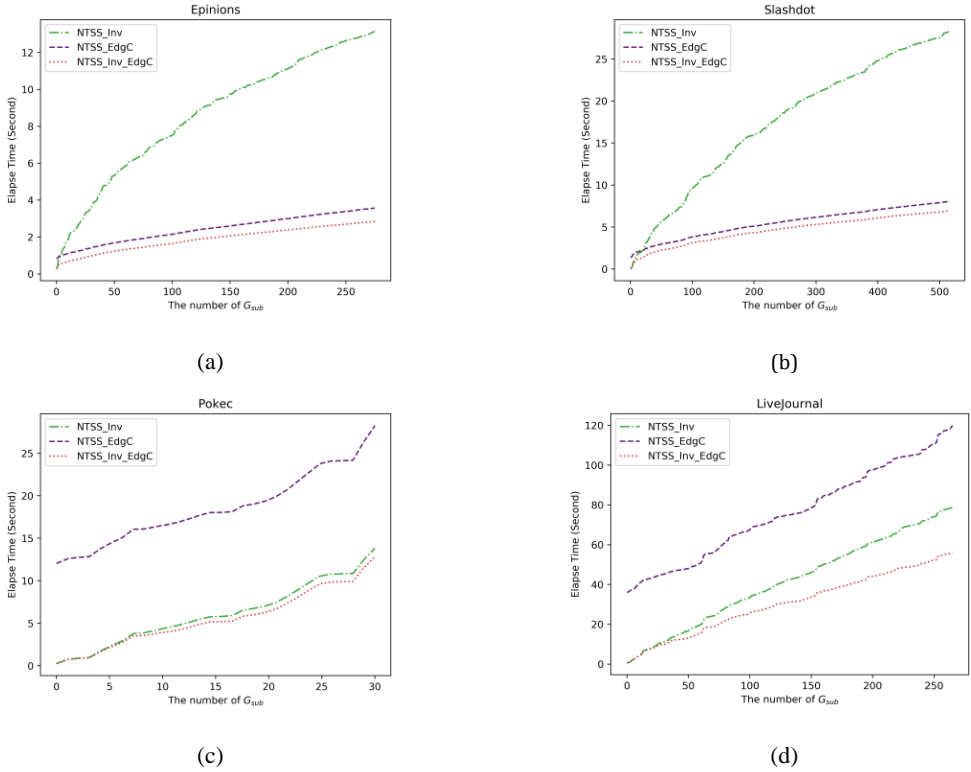

Fig. 12. The change in the number of results returned by the algorithm with optimization strategy over time when matching $P_2$

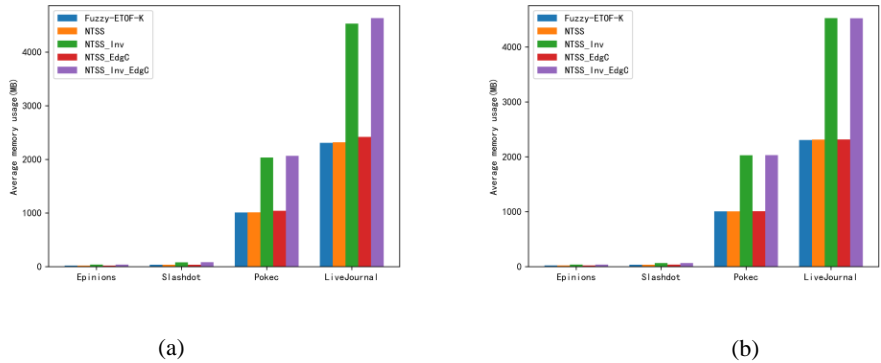

Fig. 13. Memory usage of different algorithms

**ACKNOWLEDGMENTS**

This work has been supported by the National Natural Science Foundation of China under Grant No. 62076087 & 91746209, and the Program for Changjiang Scholars and Innovative Research Team in University (PCSIRT) of the Ministry of Education under Grant No. IRT17R32.